\documentclass[usenatbib]{mn2e}
\usepackage{psfig}

\newcommand{\spose}[1]{\hbox to 0pt{#1\hss}}
\newcommand{\approxpropto}{\mathrel{\spose{\lower 3pt\hbox{$\sim$}}
	\raise 2.0pt\hbox{$\propto$}}}
\def\approxgt{\mathrel{\spose{\lower 3pt\hbox{$\sim$}}
	\raise 2.0pt\hbox{$>$}}}
\def\approxlt{\mathrel{\spose{\lower 3pt\hbox{$\sim$}}
	\raise 2.0pt\hbox{$<$}}}
\def \hkpc{\, h^{-1} \, \hbox{kpc}}
\def \hMpc{\, h^{-1} \, \hbox{Mpc}}
\def \hMsun{\, h^{-1} \, \hbox{M}_{\odot}}

\title[Entropy distribution in clusters]
{The entropy distribution in clusters: evidence of feedback?}

\author[S.~T.~Kay]
       {Scott T.~Kay\thanks{E-mail: s.t.kay@sussex.ac.uk}\\
	Department of Physics and Astronomy, University of Sussex, Falmer, 
	Brighton BN1\,9QJ}

\date{\today}

\begin{document}
\journal{Preprint astro-ph/0309435} 
 
\maketitle

\begin{abstract}
The entropy of the intracluster medium at large radii
has been shown recently to deviate from the self-similar
scaling with temperature. Using $N$-body/hydrodynamic 
simulations of the $\Lambda$CDM cosmology, we demonstrate
that this deviation is evidence that feedback processes 
are important in generating excess entropy in clusters. 
While radiative cooling increases the entropy of 
intracluster gas, resulting in a good match to the data in
the centres of clusters, it produces an entropy-temperature 
relation closer to the self-similar scaling at larger radii. 
A model that includes feedback from galaxies, however,
not only stabilises the cooling rate in the simulation, but
is capable of reproducing the observed scaling behaviour both 
in cluster cores and at large radii. Feedback modifies the
entropy distribution in clusters due to its increasing ability
at expelling gas from haloes with decreasing mass. The strength 
of the feedback required, as suggested from our simulations, 
is consistent with supernova energetics, providing a large
fraction of the energy reaches low-density regions and is
originally contained within a small mass of gas.
\end{abstract}


\section{Introduction}

Entropy\footnote{We define entropy as $S=kT/n_{\rm e}^{\gamma-1}$ keV cm$^2$,
where $\gamma=5/3$ is the ratio of specific heats for a monatomic
ideal gas.}
has become the standard quantity for describing the
distribution of the intracluster medium. In a self-similar 
population of clusters, where entropy originates solely due to 
gravitational infall via shock-heating, entropy scales 
proportional to the system's virial temperature, $S \propto T_{\rm vir}$. 
Such a model, however, does not reproduce the observed X-ray properties of
groups and clusters. The most striking example of this failure is the X-ray
luminosity-temperature ($L_{\rm X}-T_{\rm X}$) relation: 
the self-similar scaling, $L_{\rm X} \propto T_{\rm X}^{\alpha}$,
where $\alpha=2$ (Kaiser 1986) 
is flatter than the observed relation for clusters,
$\alpha \sim 3$ (e.g. Edge \& Stewart 1991;
Mushotzky \& Scharf 1997). 

This deficit in luminosity, primarily in low-mass systems, is due
to an excess of entropy in cluster cores (Evrard \& Henry 1991; 
Kaiser 1991), a phenomenon first measured by Ponman, Cannon \& Navarro 
(1999). Understanding the origin of this entropy is central to our 
understanding of cluster physics (Bower 1997; Voit et al. 2002, 2003). 
Directly heating the intracluster gas is the most common solution to this problem,
however a variety of theoretical studies conclude that the obvious
candidates -- supernovae -- are at best only marginally capable of 
generating the required excess entropy (e.g. Balogh, Babul \& Patton
1999; Kravtsov \& Yepes 2000; Loewenstein 2000; Wu, Fabian \& Nulsen 2000; 
Bower et al. 2001; Borgani et al. 2002). More energetic forms of supernovae
(so-called hypernovae) may be important. Alternative sources of
energy that are clearly observed to be interacting with the intracluster medium
are Active Galactic Nuclei (AGN), capable of distributed heating through
their release of bubbles and jets (e.g. Quilis, Bower \& Balogh 2001;
Br\"{u}ggen \& Kaiser 2002; Omma et al. 2003). 

Recently, attention has also focused on the effects of radiative 
cooling, which selectively removes the low-entropy material to form galaxies, 
causing higher-entropy material to flow in to
replace it (Knight \& Ponman 1997; Pearce et al. 2000; 
Bryan 2000; Muanwong et al. 2001, 2002; Dav\'{e}, Katz \& Weinberg 2002;
Wu \& Xue 2002). 
Cooling is an appealing mechanism as it involves no free parameters
(other than the metallicity of the gas, which is observable): the observed 
level of excess entropy is consistent with the removal of low-entropy gas 
with cooling times shorter than the age of the Universe, scaling as
$S \propto T^{2/3}$ (Voit \& Bryan 2001). 
Cooling, if unchecked however, overproduces the mass in galaxies, requiring
a feedback mechanism in order to regulate itself (White \& Rees 1978; 
Cole 1991; Balogh et al. 2001), but as was pointed out by Voit \& Bryan (2001), 
their argument still holds so long as feedback efficiently transports cold gas away from
cluster cores.

Perhaps the likely source of excess entropy is from a combination of cooling and heating
processes (Kay, Thomas \& Theuns 2003; Tornatore et al. 2003). For example, 
Kay et al. (2003) studied the combined effects of cooling and feedback 
(by reheating cold galactic gas) in simulations of groups and concluded that 
feedback, as well as regulating cooling, could contribute to the excess core entropy 
in systems with virial temperatures smaller than the heating temperature. In larger
systems where this temperature inequality was reversed, the inability of reheated gas 
to escape from the halo reduces the average entropy of the gas, and if the heating 
temperature is low enough, will affect the core entropy.

Attention has now shifted to observing the entropy of intracluster
gas at larger radii. In a recent paper, Ponman, Sanderson \& Finoguenov (2003)
measured the entropy-temperature relation at a significant fraction
of the virial radius ($R_{500}$) for a sample of 66
galaxies, groups and clusters: the Birmingham-CfA Cluster Scaling Project
(see also Sanderson et al. 2003; Sanderson \& Ponman 2003). In their study,
they measured the same deviation from self-similarity ($S \propto T^{2/3}$)
already observed in the core. This scaling behaviour was also verified recently by
Pratt \& Arnaud (2003), using {\it XMM-Newton} observations of 
Abell 1413 \& 1983. In this paper, we will argue that these observations
are evidence that feedback plays a significant role in forming the 
entropy distribution in clusters.

The rest of this paper is organized as follows. In Section~2 we
describe the simulations used to perform this study. Our 
results are presented in Section~3 and summarized in Section~4.

\section{Method}

\subsection{Simulation details}

We have performed four simulations of the $\Lambda$CDM 
cosmology, setting $\Omega_{\rm m}=0.3$, $\Omega_{\Lambda}=0.7$, 
$\Omega_{\rm b}=0.045$, $h=0.7$ \& $\sigma_8=0.9$, consistent with
the {\it WMAP} results. Each run started from the same initial
conditions: a regular cubic mesh of 2,097,152 ($128^3$) particles 
(each of gas and dark matter) within a box of comoving length 
$60 \hMpc$. This set the gas and dark matter particle masses to be 
$m_{\rm gas} \sim 1.3 \times 10^{9}$ and 
$m_{\rm CDM} \sim 7.3 \times 10^{9} \hMsun$ respectively. 
The runs were started at $z=49$ and evolved to $z=0$, using a
gravitational softening length equal to $20 \hkpc$ in comoving
co-ordinates.

The code used in this study is a parallel (MPI) version
of {\sc gadget} (Springel, Yoshida \& White 2001), an 
$N$-body/hydrodynamics code that uses a PM-tree to calculate
gravitational forces and Smooth Particle Hydrodynamics 
(SPH, e.g. Monaghan 1992) to calculate gas forces. This
version of {\sc gadget} uses entropy as the state variable
in the time-integration of the gas (Springel \& Hernquist 2002).

Differences between each of the four runs are solely due to the processes
incorporated that are able to change the entropy of the gas. These 
details are described below.

\subsubsection{Non-radiative model} 
In this run, gas was only able to increase its entropy through 
shocks. This model reproduces self-similar scalings rather well.

\subsubsection{Radiative model}
Gas in this model could also lose entropy through radiative
processes. We implemented the isochoric approximation described
in Thomas \& Couchman (1992), using equilibrium 
cooling tables from Sutherland \& Dopita (1993). We fixed the
metallicity of the gas to $Z=0.3Z_{\odot}$. Dense gas ($n>10^{-3}$cm$^{-3}, 
\delta>100$) which cooled below $T=1.2\times 10^{4}$K was converted into 
collisionless `stars'. This model significantly overproduces the 
mass fraction of cooled baryons (observed to be 5-10 per cent; Cole
et al. 2001; Balogh et al. 2001), producing a global cooled fraction of 
$\sim 34$ per cent at $z=0$.

\subsubsection{Feedback model}
To stabilize the cooling rate, we only allow some of the cooled
gas to form stars and reheat the rest of the cold material. Similarly 
to Kay et al. (2003) we incorporate feedback effects on a probabilistic
basis. (A more sophisticated heating mechanism is unwarranted for simulations
of this resolution, however this method is designed to capture the gross 
behaviour of gas being transported out of galaxies.) Rather than 
instantaneously convert all cooled particles into stars, we assign a value, 
$f_{\rm heat}$ to each cooled gas particle, defined to be the fractional 
mass of cooled material that is reheated, and draw a random number, $r$,
uniformly from the unit interval.
If $r<f_{\rm heat}$ then the particle is heated, otherwise it becomes 
collisionless. Rather than heating particles to a fixed temperature, we instead 
assigned them a fixed entropy, as it is this quantity that determines how far the 
gas can rise buoyantly out of the cluster. 

We ran two simulations with feedback, hereafter referred to as
{\it Weak Feedback} and {\it Strong Feedback} runs. In the former
case we set $f_{\rm heat}=0.5$ (i.e. a cooled particle has 
an equal chance of forming stars or being reheated) and supply an entropy  
$S_{\rm heat} = 100$ keV cm$^2$ to each reheated particle. For the
{\it Strong Feedback} simulation, we set $f_{\rm heat}=0.1$ and 
$S_{\rm heat} = 1000$ keV cm$^2$. For a cold gas particle at the same 
density, both runs require comparable amounts of energy
per mass of collisionless material but in the {\it Strong Feedback} 
case, the energy is added to one fifth 
of the mass of gas in the {\it Weak Feedback} case.

Note that an associated temperature depends on the density
of the gas. At $n=10^{-3}$cm$^{-3}$ (where most of the gas is heated in
the simulation) both runs require  $\sim 1$ keV of energy per particle, 
comparable with the energy available from supernovae. However, cold
gas within galaxies (our simulations do not resolve the internal structure
of these objects) is generally at much higher densities, where much higher
temperatures would be required to reach $S_{\rm heat}$ (if the
gas flows out adiabatically). The solution to
this problem is either that a large fraction of energy is transported
to low-density regions in kinetic form (e.g. Strickland \& Stevens 2000)
or that more energetic phenomena are at work, for example hypernovae or 
AGN or both.

\subsection{Cluster selection}

Clusters in each of the simulations were selected using the method
described in detail in Muanwong et al. (2002). A minimal-spanning tree
is first created, of all dark matter particles whose overdensity exceeds
the virial value $\delta \sim 178 \Omega^{-0.55}(z)$ (Eke, Navarro \& Frenk
1998). The tree is then pruned using a linking length, $l = 0.5\delta^{-1/3}$
times the mean interparticle separation. Spheres are then grown around
the position of maximum density in each clump of particles until the
enclosed mean density (of all particles) exceeds $\Delta$ times the
comoving critical density. In this paper we focus on results for $\Delta=500$
and $\Delta=200$. Finally, clusters with fewer than 500 particles each of
baryons and dark matter are discarded.

\section{Results}

\subsection{Entropy-Temperature relation}

\begin{figure}
\psfig{figure=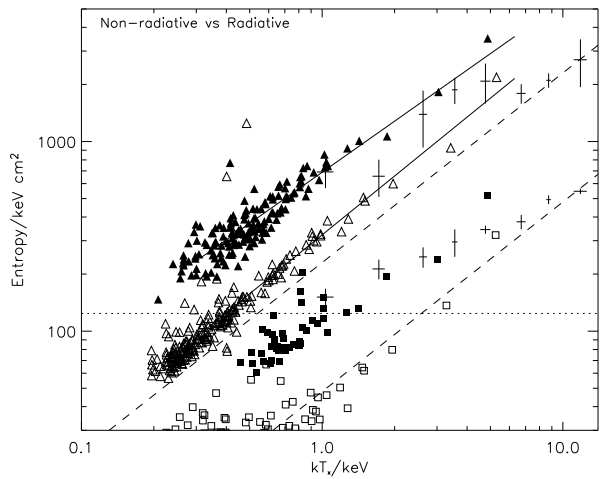,width=8.5cm}
\caption{Entropy at fixed overdensity versus 
temperature for clusters in the {\it Non-radiative}
(open symbols) and {\it Radiative} (filled symbols)
runs at $z=0$. Squares represent entropy values at $0.1R_{200}$ 
and triangles at $R_{500}$. Solid lines represent least-squares 
fits to the simulated data at $R_{500}$. The data-points are results 
from Ponman et al. (2003), with the dashed line illustrating the 
self-similar slope ($S \propto T$), normalized to their 8 hottest 
clusters. The horizontal dotted line is the entropy floor, 
$S \sim 124$ keV cm$^2$, from Lloyd-Davies et al. (2000).}
\label{fig:strel1}
\end{figure}

We begin by showing results from our {\it Non-radiative} and
{\it Radiative} simulations. Fig.~\ref{fig:strel1} illustrates
entropy at a fixed density contrast against soft-band X-ray 
temperature for clusters in these two runs. The temperatures
are calculated as
\begin{equation}
T_{\rm X} = \frac{\Sigma_i m_i\rho_i\Lambda_{\rm soft}(T_i,Z)T_i}
              {\Sigma_i m_i\rho_i\Lambda_{\rm soft}(T_i,Z)},
\label{eq_9}
\end{equation}
where $m_i$, $\rho_i$ and $T_i$ are the mass, density and temperature
of hot gas particle $i$ ($T_i>10^{5}$K) and $Z=0.3Z_{\odot}$. 
We adopt the soft-band cooling function, $\Lambda_{\rm soft}$, 
from Raymond \& Smith (1977) for an energy range 0.3--1.5\,keV. 
We calculate a mass-weighted entropy for each cluster by averaging
the entropy of hot gas particles within a shell of width 20$\hkpc$ 
(our results are insensitive to changing the size of this shell by a 
factor of 2), centred on the radius of interest. 

Squares in the figure represent entropy
values at $0.1R_{200}$, a radius typically chosen to highlight
excess entropy in groups (Ponman et al. 1999). As
shown previously (Muanwong et al. 2002; Kay et al. 2003), 
the {\it Non-radiative} model (open symbols) agrees well with
the self-similar prediction ($S \propto T$), apart from at low temperatures
where there is some scatter due to insufficient resolution in the core,
while the {\it Radiative} model (filled symbols) is in reasonable agreement 
with the observations (the error-bars shown here are from Ponman et al. 2003).

\begin{figure}
\psfig{figure=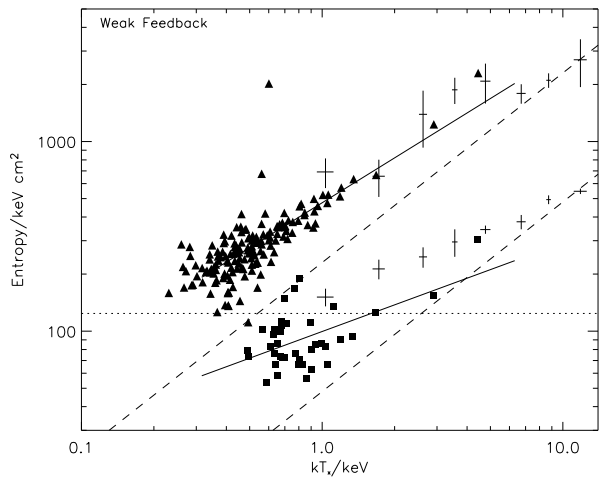,width=8.5cm}
\psfig{figure=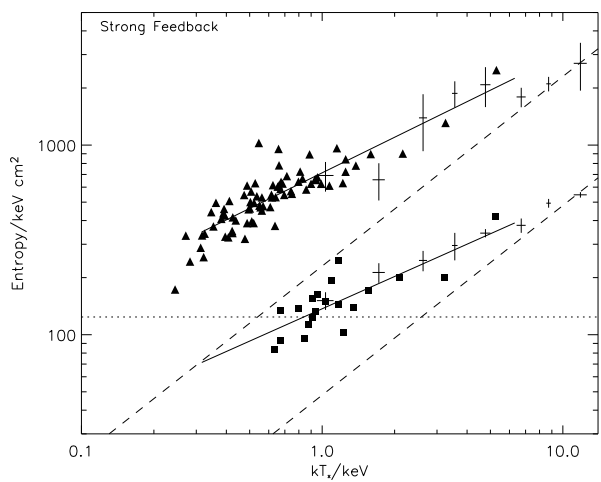,width=8.5cm}
\caption{Entropy at fixed overdensity versus 
temperature for clusters in the {\it Weak Feedback} 
(top panel) and {\it Strong Feedback} (bottom panel) runs
at $z=0$. Squares represent entropy values at $0.1R_{200}$ 
and triangles at $R_{500}$. Solid lines represent least-squares 
fits to the simulated data. The data-points are results from 
Ponman et al. (2003), with the dashed line illustrating the 
self-similar slope ($S \propto T$), normalized to their 8 
hottest clusters. The horizontal dotted line is the entropy 
floor, $S \sim 124$ keV cm$^2$, from Lloyd-Davies et al. (2000).}
\label{fig:strel2}
\end{figure}

We also show in this figure (triangles) results for a 
lower density contrast of 500, corresponding to $\sim 2/3 R_{200}$. 
Again, our {\it Non-radiative} results agree well with the self-similar
scaling but (allowing extrapolation) are around 40 per cent higher than 
the expected result when normalized to the hottest 8 clusters in the Ponman et al. sample 
(upper dashed line). Note however, that the {\it Radiative} results at $R_{500}$
are higher than the observational data at all temperatures, but scale 
almost self-similarly ($S \propto T_{\rm X}^{0.9}$).
Cooling increases the entropy in clusters at all radii but in this case, 
disagrees with the observed slope. It is therefore apparent that cooling 
has been too efficient at generating excess entropy in the most 
massive clusters.

Fig.~\ref{fig:strel2} illustrates the same relationships for our
{\it Weak Feedback} and {\it Strong Feedback} runs. In the former
case, we see that this model does a reasonable job at $R_{500}$,
flatter than the self-similar relation, but the entropy at $0.1R_{200}$ 
is considerably lower than observed. In this model, the gas is not 
receiving enough entropy to escape from the inner (X-ray dominant) 
region of the halo, but the feedback is having a larger effect on
the gas in lower-mass systems, even at $R_{500}$. (Note that reheated
gas can lose some of its entropy through cooling.)

The {\it Strong Feedback} model, however, matches both relations very 
well. The increase in entropy allows the gas to escape to larger radii
than previously, but the fate of this gas depends on the system
size. In low-temperature systems, a significant fraction of hot gas 
gains enough entropy to escape the system entirely (Kay et al. 2003), 
resulting in an entropy excess that is shown here to be present even
at large radii. This effect diminishes in higher-temperature systems,
where the entropy level of the gas is sufficiently high that less
material was able to escape.

\subsection{Profiles}

\begin{figure}
\psfig{figure=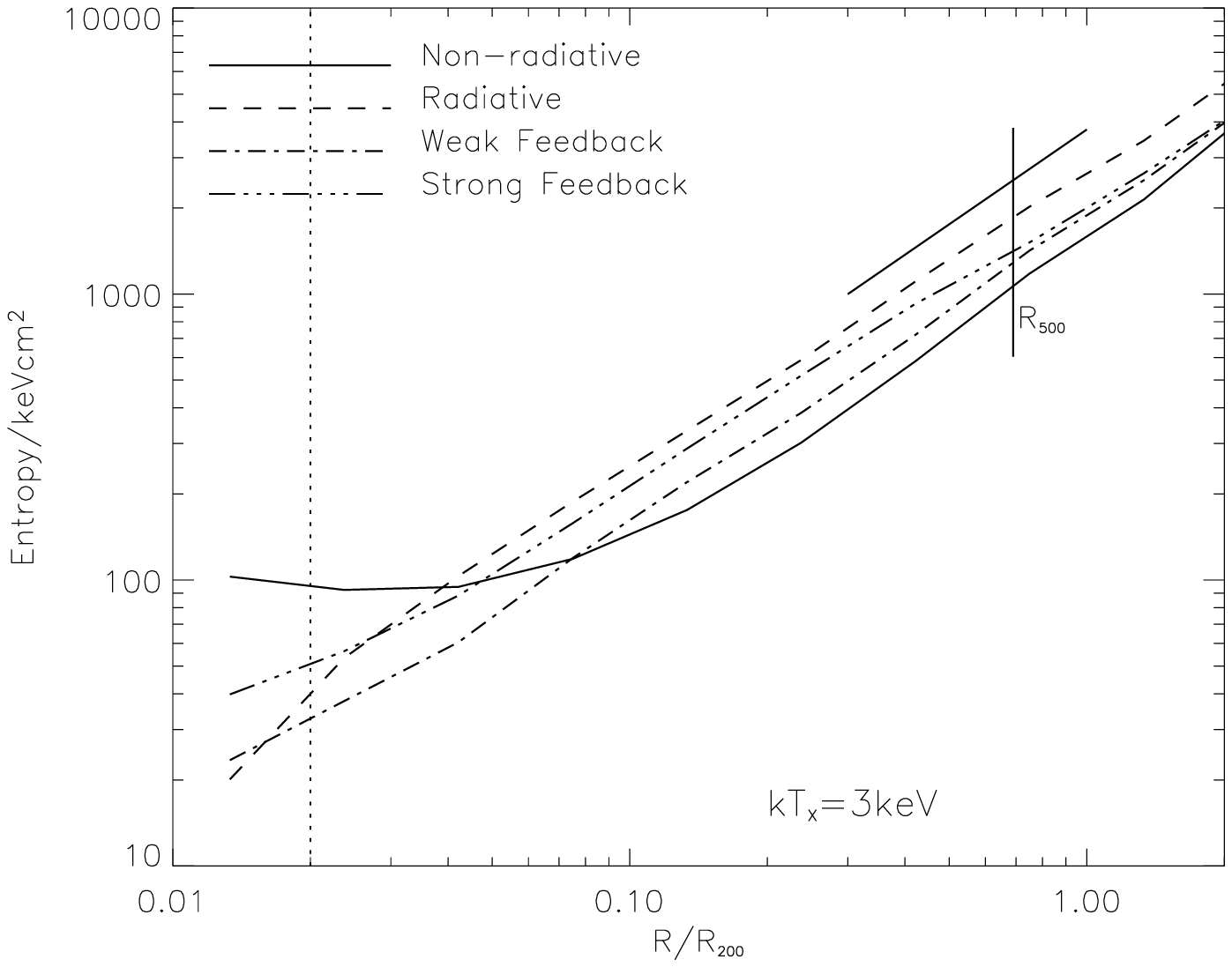,width=8.7cm}
\psfig{figure=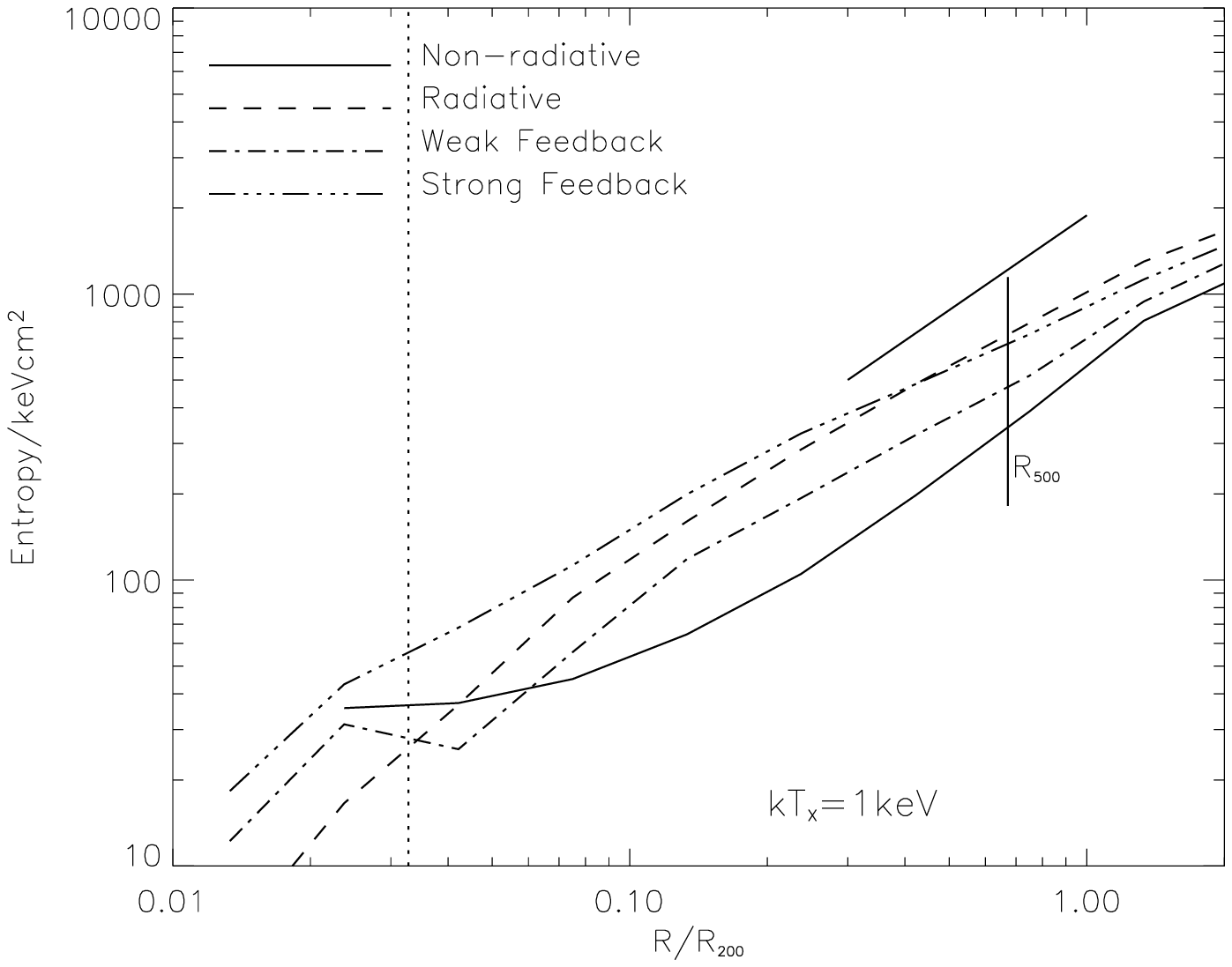,width=8.7cm}
\caption{Entropy profiles for two haloes (top panel:
$kT_{\rm X}\sim 3$ keV and bottom panel: $kT_{\rm X} \sim 1$ keV)
at $z=0$, from the {\it Non-radiative} simulation (solid curve),
{\it Radiative} simulation (dashed curve), {\it Weak Feedback} 
simulation (dot-dashed curve) and the {\it Strong Feedback} 
simulation (triple-dot-dashed curve). The vertical dotted
line marks the softening radius and the solid vertical line 
$R_{500}$. The solid diagonal line illustrates the predicted
slope from spherical accretion shock models, $S \propto R^{1.1}$.}
\label{fig:entprof}
\end{figure}

To illustrate more clearly the effects of cooling and heating on the
entropy distribution of our clusters, we plot in Fig.~\ref{fig:entprof}
entropy profiles for two systems with $kT_{\rm X} \sim 1$ \& 3 
keV respectively. At large radii in both cases, the {\it Non-Radiative} model 
has an entropy profile in reasonable agreement with the relation, 
$S \propto R^{1.1}$, as expected from spherical accretion shocks
(Tozzi \& Norman 2001). Both profiles flatten within $\sim $10 per cent
of $R_{200}$.

Cooling increases the entropy of the gas at all radii (by removing
a significant fraction of gas over the age of the universe) except for 
in the very centre where the cooling time is very short, causing
the entropy to drop sharply. These effects are more prominent in
the smaller system, reflecting the increasing efficiency of 
cooling with decreasing halo mass. 

The {\it Weak Feedback} profiles are similar to the {\it Radiative} 
profiles, except that they are lower in normalization. The feedback 
in this case is enough to prevent some gas from cooling but not to move 
it around in the cluster significantly. The {\it Strong feedback} model  
on the other hand shows greater differences. In the 1 keV system 
the entropy profile is above the {\it Radiative} profile out to 
$\sim 0.4 R_{200}$, where they both become comparable. Feedback therefore has
increased the core entropy over and above what is possible from cooling
alone, by moving some of the hot gas (which did not cool in the 
{\it Radiative} model) to larger radii.
In the hotter system, however, the {\it Strong Feedback} 
profile is slightly lower than the {\it Radiative} profile out to 
$\sim 0.2 R_{200}$, where it decreases further, and at $R_{500}$ is only 
slightly higher than the {\it Non-radiative} profile. In this system,
a larger fraction of gas was retained by the halo due to its deeper
gravitational potential well.

We note that the {\it Strong Feedback} profiles are a bit flatter 
than $R^{1.1}$ at $R_{500}$ (supported by observations). Generally, 
systems above 1 keV have slopes between 0.8-0.9. Whether this
discrepancy is a problem with this model requires further investigation,
but will greatly benefit from both larger simulations and a larger 
sample of high-quality cluster data.

\section{Summary}

In this paper we used results from $N$-body/hydrodynamic
simulations of the $\Lambda$CDM cosmology with various degrees
of cooling and feedback, in order to understand the recent
observational claim (Ponman et al. 2003; Pratt \& Arnaud 2003)
that the entropy of intracluster gas does not scale self-similarly 
with temperature out to at least $R_{500}$ ($\sim 2/3 R_{200}$). 
In particular, we compared our results to the entropy-temperature
relation found by Ponman et al. at $0.1R_{200}$ and at $R_{500}$,
where both are found to be flatter ($S \propto T^{\sim 2/3}$) than the 
self-similar scaling ($S \propto T$). Our results can be summarized 
as follows.

\begin{itemize}
\item Non-radiative clusters scale self-similarly in
the entropy-temperature plane to good approximation, both at
$0.1R_{200}$ and at $R_{500}$. All but the largest
clusters are observed to have an excess of entropy relative
to the self-similar model at both radii.

\item Radiative cooling raises the entropy in clusters at all radii.
While this reproduces the entropy-temperature relation well at
$0.1R_{200}$, it produces a relation that is too steep at $R_{500}$.
In this case, the relation is almost self-similar, suggesting that 
clusters above a few keV have too much entropy compared to observations.
This is due to the hot gas mass being too low at all radii.

\item A feedback model in which gas receives a sufficient
(1000 keV cm$^2$) amount of entropy results in excellent agreement
with the observations at both radii. The reheated gas has enough 
entropy to escape the X-ray core but the fraction of material 
unable to leave the cluster (particularly at large radii) 
increases with system size, flattening the entropy-temperature 
relation. In detail, the entropy profiles are a bit flatter (at $R_{500}$) 
than suggested by observations, prompting future investigation into
whether this discrepancy is significant.

\end{itemize}

We conclude that the entropy level required to match the observations
is high, but supernovae are energetically capable, providing the energy is
efficiently transported into lower density regions and contained
within a small mass of gas. Alternatively, more energetic phenomena 
(hypernovae or Active Galactic Nuclei) may play a part.

\section*{Acknowledgements}
We thank Peter Thomas for many helpful discussions and comments on the 
original manuscript, Trevor Ponman for providing useful comments
and supplying data-points from the Birmingham-CfA Cluster Scaling Project
and Volker Springel for making his code {\sc gadget} available to us.
The simulations used in this paper were carried out using the Cosmology
Machine at the Institute for Computational Cosmology, Durham as part 
of the Virgo Consortium programme of investigations into the formation 
of structure in the Universe. STK is supported by PPARC.


\begin{thebibliography}{}

\bibitem[\protect\citeauthoryear{Balogh, Babul \& Patton }1999]{balogh99}
	Balogh M.~L., Babul A., Patton D.~R., 1999, MNRAS, 307, 463
\bibitem[\protect\citeauthoryear{Balogh et al. }2001]{balogh01} 
	Balogh M.~L., Pearce F.~R., Bower R.~G., Kay S.~T., 2001,
	MNRAS, 326, 1228
\bibitem[\protect\citeauthoryear{Bower }1997]{bower97} Bower R. G.,
	1997, MNRAS, 288, 355
\bibitem[\protect\citeauthoryear{Bower et al. }2001]{bower01} 
	Bower R. G., Benson A. J., Lacey C. G., Baugh C. M., Cole S.,
	Frenk C. S., 2001, MNRAS, 325, 497
\bibitem[\protect\citeauthoryear{Br\"{u}ggen \& Kaiser }2002]{bruggen02}
	Br\"{u}ggen M., Kaiser C.~R., 2002, Nature, 418, 301
\bibitem[\protect\citeauthoryear{Bryan }2000]{bryan00}
	Bryan G. L., 2000, ApJ, 544, L1
\bibitem[\protect\citeauthoryear{Cole }1991]{cole91}
	Cole S., 1991, ApJ, 367, 45
\bibitem[\protect\citeauthoryear{Cole et al. }2001]{cole01}
	Cole S. et al., 2001, MNRAS, 326, 255
\bibitem[\protect\citeauthoryear{Dav\'{e}, Katz \& Weinberg }2002]{dave02}
	Dav\'{e} R., Katz N., Weinberg D. H., 2002, ApJ, 579, 23
\bibitem[\protect\citeauthoryear{Edge \& Stewart }1991]{es91} Edge A. C., 
        Stewart G. C., 1991, MNRAS, 252, 414
\bibitem[\protect\citeauthoryear{Eke, Navarro \& Frenk }1998]{eke98}
	Eke V. R., Navarro J. F., Frenk C. S., 1998, ApJ, 503, 569
\bibitem[\protect\citeauthoryear{Evrard \& Henry }1991]{evrard91}
	Evrard A. E., Henry J. P., 1991, ApJ, 383, 95
\bibitem[\protect\citeauthoryear{Kaiser }1986]{k86} Kaiser N., 1986, MNRAS, 
        222, 323
\bibitem[\protect\citeauthoryear{Kaiser }1991]{k91} Kaiser N., 1991, ApJ,
	383, 104
\bibitem[\protect\citeauthoryear{Kay, Thomas \& Theuns }2003]{kay03}
	Kay S.~T., Thomas P.~A., Theuns T., 2003, MNRAS, 343, 608
\bibitem[\protect\citeauthoryear{Knight \& Ponman }1997]{knight97}
	Knight P. A., Ponman T. J., 1997, MNRAS, 289, 955
\bibitem[\protect\citeauthoryear{Kravtsov \& Yepes }2000]{kravtsov00} 
	Kravtsov A.~V., Yepes G., 2000, MNRAS, 318, 227
\bibitem[\protect\citeauthoryear{Lloyd--Davies, Ponman \& Cannon }2000]{ld2000}
	Lloyd--Davies E.~J., Ponman T.~J., Cannon D.~B., 2000, MNRAS, 315, 689
\bibitem[\protect\citeauthoryear{Loewenstein }2000]{loewenstein00}
	Loewenstein M., 2000, ApJ, 532, 16
\bibitem[\protect\citeauthoryear{Monaghan }1992]{MON} Monaghan J.J., 1992, ARA\&A, 30, 543
\bibitem[\protect\citeauthoryear{Muanwong et al. }2001]{muanwong01} Muanwong 
        O., Thomas P. A., Kay S. T., Pearce F. R., Couchman H. M. P., 2001, 
        MNRAS, 552, L27
\bibitem[\protect\citeauthoryear{Muanwong et al. }2002]{muanwong02} Muanwong 
        O., Thomas P. A., Kay S. T., Pearce F. R., 2002, MNRAS, 336, 527
\bibitem[\protect\citeauthoryear{Mushotzky \& Scharf }1997]{mushotzky97}
	Mushotzky R.~F., Scharf C.~A., 1997, ApJ, 482, L13
\bibitem[\protect\citeauthoryear{Navarro, Frenk \& White }1995]{nfw95}
        Navarro J. F., Frenk C. S., White S. D. M., 1995, MNRAS, 275, 720
\bibitem[\protect\citeauthoryear{Omma et al. }2003]{omma03}
	Omma H., Binney J., Bryan G., Slyz A., 2003, MNRAS, submitted 
	(astro-ph/0307471)
\bibitem[\protect\citeauthoryear{Pearce et al. }2000]{pearce00}
	Pearce F.~R., Thomas P.~A., Couchman H.~M.~P., Edge A.~C., 2000, 
	MNRAS, 317, 1029
\bibitem[\protect\citeauthoryear{Ponman, Cannon \& Navarro }1999]{ponman99}
	Ponman T.~J., Cannon D.~B., Navarro  J.~F., 1999, Nature, 397, 135
\bibitem[\protect\citeauthoryear{Ponman, Sanderson \& Finoguenov 
	}2003]{ponman03} Ponman T. J., Sanderson A. J. R., Finoguenov A., 
	2003, MNRAS, 343, 331 
\bibitem[\protect\citeauthoryear{Pratt \& Arnaud }2003]{pratt03}
	Pratt G.~W., Arnaud M., 2003, A\&A, 408, 1
\bibitem[\protect\citeauthoryear{Quilis, Bower \& Balogh }2001]{quilis01}
	Quilis V., Bower R. G., Balogh M. L., 2001, MNRAS, 328, 1091
\bibitem[\protect\citeauthoryear{Raymond \& Smith }1977]{rs77}
	Raymond J. C., Smith B. W., 1977, ApJS, 35, 419
\bibitem[\protect\citeauthoryear{Sanderson et al. }2003]{sanderson03a}
	Sanderson A.~J.~R., Ponman T.~J., Finoguenov A., Lloyd-Davies E.~J.,
	Markevitch M., 2003, MNRAS, 340, 989
\bibitem[\protect\citeauthoryear{Sanderson \& Ponman }2003]{sanderson03b}
	Sanderson A.~J.~R., Ponman T.~J., 2003, MNRAS, accepted 
	(astro-ph/0307457)
\bibitem[\protect\citeauthoryear{Springel, Yoshida \& Hernquist }2001]
	{springel01} Springel V., Yoshida N., White S.~D.~M., 2001, New Astronomy,
	6, 79
\bibitem[\protect\citeauthoryear{Springel \& Hernquist }2002]
	{springel02} Springel V., Hernquist L., 2002, MNRAS, 333, 649
\bibitem[\protect\citeauthoryear{Strickland \& Stevens }2000]
	{strickland02} Strickland D.~K., Stevens I.~R., 2000, MNRAS, 314, 511
\bibitem[\protect\citeauthoryear{Sutherland \& Dopita }1993]{sd93}
	Sutherland R. S., Dopita M. A., 1993, ApJS, 88, 253
\bibitem[\protect\citeauthoryear{Tornatore et al. }2003]{tornatore03}
	Tornatore L., Borgani S., Springel V., Matteucci F., Menci N.,
	Murante G., 2003, MNRAS, 342, 1025
\bibitem[\protect\citeauthoryear{Tozzi \& Norman }2001]{tozzi01}
	Tozzi P., Norman C., 2001, ApJ, 546, 63
\bibitem[\protect\citeauthoryear{Voit \& Bryan }2001]{voit01} 
	Voit G.~M., Bryan G.~L., 2001, Nature, 414, 425
\bibitem[\protect\citeauthoryear{Voit et al. }2002]{voit02}
	Voit G. M., Bryan G. L., Balogh M. L., Bower R. G., 2002,
	ApJ, 576, 601
\bibitem[\protect\citeauthoryear{Voit et al }2003]{voit03}
	Voit G.~M., Balogh M.~L., Bower R.~G., Lacey C.~G.,
	Bryan G.~L., 2003, ApJ, 593, 272
\bibitem[\protect\citeauthoryear{White \& Rees }1978]{white78} 
	White S.~D.~M.,  Rees M.~J., 1978, MNRAS, 183, 341
\bibitem[\protect\citeauthoryear{Wu, Fabian \& Nulsen }2000]{wu00} 
	Wu K.~K.~S., Fabian A.~C., Nulsen P.~E.~J., 2000, MNRAS, 318, 889
\bibitem[\protect\citeauthoryear{Wu \& Xue }2002]{wu02}
	Wu X.-P., Xue Y.-J., 2002, ApJ, 569, 112
\end{thebibliography}
\end{document}